\documentclass[sigconf]{acmart}

\renewcommand\footnotetextcopyrightpermission[1]{} 
\usepackage{booktabs} 
\usepackage{url}
\usepackage{mathrsfs}
\usepackage[mathscr]{eucal}
\usepackage[linesnumbered,ruled,vlined]{algorithm2e}
\usepackage{multirow}
\usepackage{subcaption}
\usepackage{amsthm}
\usepackage{amsmath}
\usepackage{graphicx}
\usepackage{array}
\usepackage{makecell}
\usepackage{setspace}

\usepackage{tikz}
\usetikzlibrary{arrows,positioning,automata,calc,shapes}
\usepackage{pgfplots}
\pgfplotsset{compat=newest, scaled z ticks=false} 
\pgfplotsset{plot coordinates/math parser=false}
\newlength\figureheight 
 \newlength\figurewidth

\newcommand{\squishlist}{
    \begin{list}{$\bullet$}
    { \setlength{\itemsep}{0pt}
        \setlength{\parsep}{1pt}
        \setlength{\topsep}{1pt}
        \setlength{\partopsep}{0pt}
        \setlength{\leftmargin}{1em} 
        \setlength{\labelwidth}{1em}
        \setlength{\labelsep}{0.5em}
    						 } }

\newcommand{\squishlisttwo}{
    \begin{list}{$\bullet$}
        { \setlength{\itemsep}{0pt}
            \setlength{\parsep}{0pt}
            \setlength{\topsep}{0pt}
            \setlength{\partopsep}{0pt}
            \setlength{\leftmargin}{2em}
            \setlength{\labelwidth}{1.5em}
            \setlength{\labelsep}{0.5em} } }

\newcommand{\squishend}{
    \end{list}  }

\begin{document}

\fancyhead{}
\title{Evolution of Popularity Bias: Empirical Study and Debiasing}

\author{Ziwei Zhu}
\authornote{Work was done when Ziwei, Yun, and Xing were at Texas A\&M University.}
\affiliation{
 \institution{George Mason University}
 \country{}}

\author{Yun He}
\affiliation{
 \institution{Texas A\&M University}
 \country{}}
   
\author{Xing Zhao}
\affiliation{
 \institution{Texas A\&M University}
 \country{}}
   
   \author{James Caverlee}
\affiliation{
 \institution{Texas A\&M University}
 \country{}}

\begin{abstract}
Popularity bias is a long-standing challenge in recommender systems. Such a bias exerts detrimental impact on both users and item providers, and many efforts have been dedicated to studying and solving such a bias. However, most existing works situate this problem in a static setting, where the bias is analyzed only for a single round of recommendation with logged data. These works fail to take account of the dynamic nature of real-world recommendation process, leaving several important research questions unanswered: how does the popularity bias evolve in a dynamic scenario? what are the impacts of unique factors in a dynamic recommendation process on the bias? and how to debias in this long-term dynamic process? In this work, we aim to tackle these research gaps. Concretely, we conduct an empirical study by simulation experiments to analyze popularity bias in the dynamic scenario and propose a dynamic debiasing strategy and a novel False Positive Correction method utilizing false positive signals to debias, which show effective performance in extensive experiments.

\end{abstract}




\maketitle

\section{Introduction}
\label{sec:intro}

Popularity bias -- popular items are overly exposed in recommendations at the expense of less popular items that users may find interesting -- is a long-standing challenge in recommender systems~\cite{wei2020model,steck2019collaborative,steck2011item,zhu2021popularity,park2008long}. Most existing efforts to study popularity bias adopt a static setting~\cite{wei2020model,steck2019collaborative,steck2011item,zhu2021popularity,park2008long}. That is, a model is trained over an offline dataset, and popularity bias is analyzed by conducting a single round of recommendation. While these studies have highlighted the prevalence of popularity bias, there is a significant research gap in our understanding of the dynamics of this bias, the factors impacting popularity bias and its evolution, and the efficacy of methods to mitigate this bias under real-world assumptions of system evolution. Hence, this paper proposes a framework for the study of popularity bias in dynamic recommendation.

\begin{figure}[t!] 
\centering
\includegraphics[ width=0.7\linewidth ]{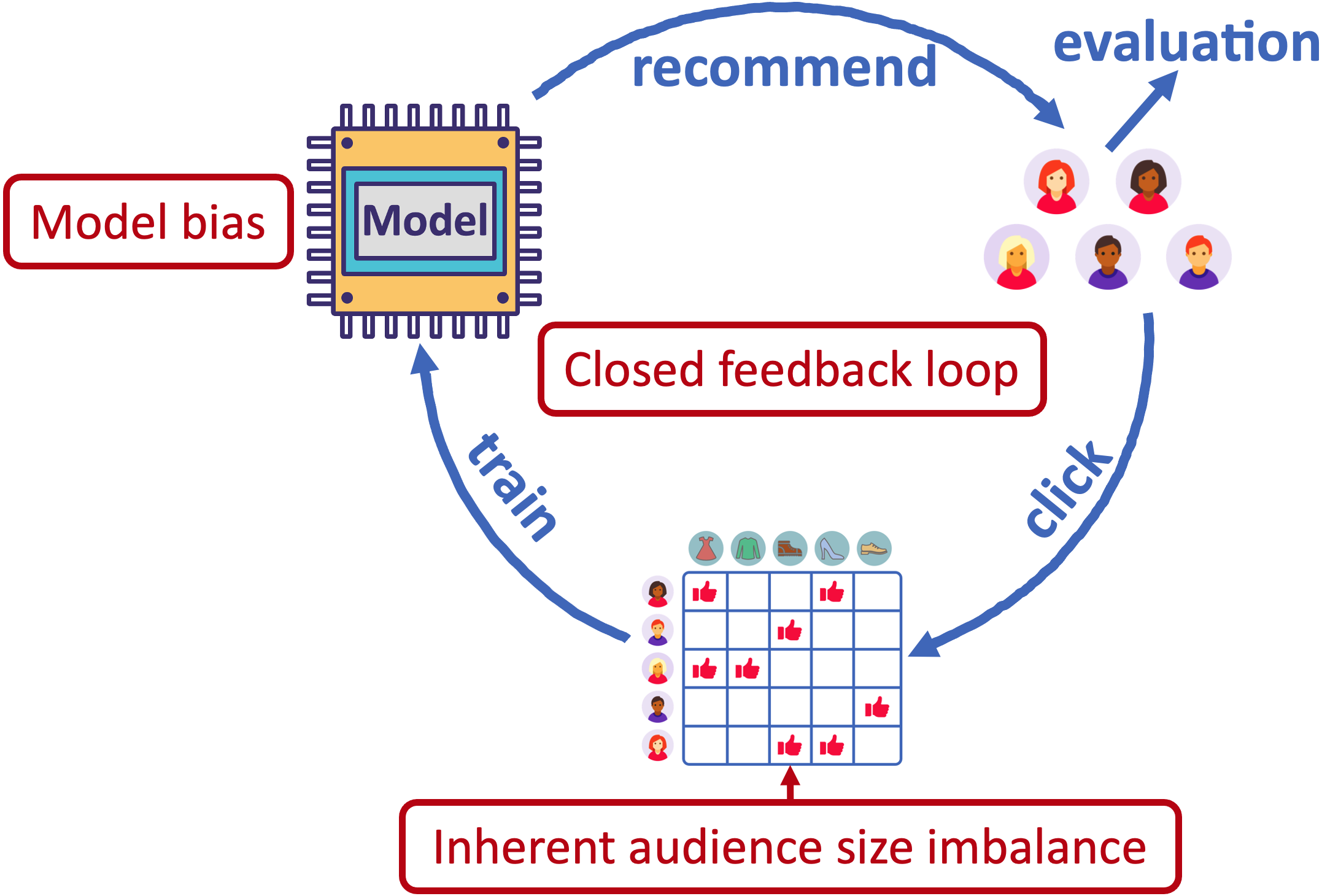} 
\vspace{-10pt} 
\caption{The pipeline of the dynamic recommendation.}
\label{fig:intro} 
\vspace{-17pt} 
\end{figure}

\textit{Dynamic recommendation}~\cite{khenissi2020theoretical,morik2020controlling,chaney2018algorithmic,sun2019debiasing} can be viewed as a closed loop illustrated in Figure~\ref{fig:intro}. Users interact with the system through a set of actions (e.g., clicks, views, ratings); this user-feedback data is then used to train a recommendation model; the trained model is used to recommend new items to users; and then the loop continues. While there are many opportunities for bias to affect this dynamic recommendation process, we identify three key factors that may impact popularity bias and its evolution: (i) \textit{inherent audience size imbalance}: users may like some items more than others (even with a purely bias-free random recommender), meaning that a few items may have very large audience sizes while the majority have small ones; (ii) \textit{model bias}: the recommendation model itself may amplify any imbalances in the data it ingests for training; and (iii) \textit{closed feedback loop}: since the cycle repeats, the feedback collected from recommendations by the current model will impact the training of future versions of the model, potentially accumulating the bias. 


Specifically, we undertake a three-part study to investigate popularity bias in dynamic recommendation: (i) We conduct a comprehensive empirical study by simulation experiments to investigate how the popularity bias evolves in dynamic recommendation, and how the three factors impact the bias. (ii) We explore methods to mitigate popularity bias in dynamic recommendation. We show how to adapt existing debiasing methods proposed in a static setting to the dynamic scenario. We further propose a model-agnostic False Positive Correction (FPC) method for debiasing, which can be integrated with other debiasing methods for further performance improvements. (iii) Finally, we report on extensive experiments to show the effectiveness of the proposed debiasing method compared with state-of-the-art baselines.


\section{Related Work}
\label{sec:related_work}

Popularity bias is a long-standing problem and has been widely studied. Some methods adopt an in-processing strategy to mitigate bias by modifying the model itself  ~\cite{abdollahpouri2017controlling,wei2020model,steck2011item}, while others adopt a post-processing strategy to mitigate bias by modifying the predictions of the model ~\cite{zhu2021popularity,steck2019collaborative,abdollahpouri2019managing}. One of the most typical approaches is to debias by assigning weights inversely proportional to item popularity in the loss of a model~\cite{steck2011item,liang2016causal}. By this, popular and unpopular items can be balanced during training and more even recommendations can be generated. Similar to this idea, Steck~\cite{steck2019collaborative} proposes to directly re-scale the predicted scores based on popularity to promote unpopular items and prevent popular items from over-recommendation. The scaling weights are also inversely proportional to item popularity. Besides, a recent work~\cite{wei2020model} investigates the bias from the perspective of causal inference and propose a counterfactual reasoning method to debias. Note that all of these works evaluate popularity bias by comparing how often items are recommended without regard for the ground truth of user-item matching. To address this gap, \cite{zhu2021popularity} proposes the concept of popularity-opportunity bias which compares the true positive rate of items to evaluate the bias, and a popularity compensation method is proposed, which explicitly considers user-item matching.


\section{Problem Formalization}

\subsection{Formalizing Dynamic Recommendation}
\label{sec:formalize_dynamic_rec}


\begin{algorithm}[t!]
\textbf{Bootstrap:} Randomly show $K$ items to each user and collect initial clicks $\mathcal{D}$\, and train the first model $\psi$ by $\mathcal{D}$\;
\For{$t=1:T$}{
    Recommend $K$ items to the current user $u_t$ by $\psi$\;
    Collect new clicks and add them to $\mathcal{D}$\;
    \If{$t\%L==0$}{
        Retrain $\psi$ by $\mathcal{D}$\;
    }
}
\label{alg:dynamic_rec} 
\caption{Dynamic Recommendation Process}
\end{algorithm}

\setlength{\textfloatsep}{2pt}

Suppose we have an online platform that provides recommendations. Given we have a set of users $\mathcal{U} = \{1, 2, \ldots, N\}$ and a set of items $\mathcal{I} = \{1, 2, \ldots, M\}$ in the system. Every user has a subset of items the user likes (unknown to the system), and we define the total number of matched users who like the item $i$ as the audience size of $i$, denoted as $A_i$. At the beginning (a bootstrap step), for each user, the system randomly exposes $K$ items to bootstrap the user and thus collects initial user-item clicks $\mathcal{D}$. Based on the initial data $\mathcal{D}$, the first recommendation model $\psi$, such as a matrix factorization (MF)~\cite{koren2009matrix}, is trained. Then, as users coming to the system one by one, the system uses the up-to-date model to provide $K$ ranked items as recommendations and collect new user-item clicks. After every $L$ user visits, the system retrains the recommendation model with all clicks collected up to now. This dynamic recommendation process is summarized in Algorithm 1.

\subsection{Formalizing Popularity Bias}
\label{sec:formalizing_bias}

We adopt the recently introduced popularity-opportunity bias~\cite{zhu2021popularity}, which evaluates \textit{whether popular and unpopular items receive clicks (or other engagement metrics) proportional to their true audience sizes}. In other words, do popular and unpopular items receive similar \textit{true positive rates}? At iteration $t$ in the dynamic recommendation process, to quantify the popularity bias, we need to first calculate the true positive rate for each item. Suppose item $i$ has received $C^t_i$ clicks in total from the beginning to iteration $t$, the true positive rate for $i$ is $TPR_i=C^t_i/A_i$. Then, we can use the Gini Coefficient~\cite{wagstaff1991measurement,brown1994using} to measure the inequality in true positive rates corresponding to item popularity at iteration $t$:
\begin{equation}
\begin{aligned}
\centering
    Gini_t = \frac{\sum_{i\in\mathcal{I}}(2i-M-1)TPR_i}{M\sum_{i\in\mathcal{I}}TPR_i},
\end{aligned}
\label{equ:GC}
\end{equation}
where items are indexed from $1$ to $M$ in audience size non-descending order ($A_i\leq A_{(i+1)}$). We use $-1\leq Gini_t\leq 1$ to quantify the popularity bias~\footnote{In this paper, we conduct simulation experiments with semi-synthetic data to study the popularity bias in dynamic recommendation, in which audience size of items are known. In practice, we need to estimate the audience size based on observed clicks, such as inverse propensity scoring based methods from \cite{morik2020controlling,yang2018unbiased}.}: a small $|Gini_t|$ indicates a low bias; $Gini_t>0$ represents that true positive rate is positively correlated to item audience size; and $Gini_t<0$ represents that the true positive rate is negatively correlated to audience size (reversed popularity bias).

\subsection{Factors Impacting Popularity Bias}
\label{sec:factors}
One of the goals in this paper is to deepen our understanding of factors that may produce and worsen this bias. As introduced in Figure~\ref{fig:intro}, we focus on three major factors:

\smallskip
\noindent\textbf{1. Inherent audience size imbalance.} Items inherently have different audience sizes, and this imbalance can potentially lead to popularity bias. It has been observed that the audience size for items usually follows a long-tail distribution~\cite{park2008long}, meaning that a few items have a very large audience size while the majority have small ones. This inherent imbalance will result in imbalanced engagement data (like clicks), even if every item is equally recommended by a bias-free random recommender.


\smallskip
\noindent\textbf{2. Model bias.} A recommendation model tends to rank an item with more clicks in the training data higher than an item with fewer clicks, even if the ground truth is that the user equally likes both of them~\cite{zhu2021popularity}. This is a common deficiency of collaborative filtering based algorithms and directly leads to popularity bias if the training data is imbalanced. 

\smallskip
\noindent\textbf{3. Closed feedback loop.} Finally, we consider the phenomenon that future models are trained by the click data collected from the recommendations by previous models~\cite{jiang2019degenerate,sinha2016deconvolving,sun2019debiasing}. In this way, the popularity bias generated in the past can accumulate, leading to more bias in subsequent models as the feedback loop continues.



\section{Empirical Study}
\label{sec:DDS}
In this section, we conduct an empirical study to uncover how the popularity bias evolves in dynamic recommendation; and the impacts of the three discussed bias factors on the bias.

\subsection{Setup}
\label{sec:DDS_setup}

\begin{table}[t!]\small
\caption {Dataset statistics.}
\vspace{-10pt}
\centering
  \renewcommand{\arraystretch}{.9}
\begin{tabular}{c|cccc}
\hline\hline
     & \#user & \#item & density & $Gini$(audience size) \\ \hline
ML1M & 1,000  & 3,406  & 0.0657  & 0.6394            \\
Ciao & 1,000  & 2,410  & 0.0696  & 0.4444            \\ \hline\hline
\end{tabular}
\label{table:datasets}
\end{table}

Due to the challenges of running repeatable experiments over live platforms, we follow the widely-adopted approach~\cite{khenissi2020theoretical,morik2020controlling,chaney2018algorithmic,aridor2020deconstructing,jiang2019degenerate} of conducting experiments to simulate the dynamic recommendation process in Section~\ref{sec:formalize_dynamic_rec}.


First, we follow \cite{morik2020controlling,khenissi2020theoretical} to generate semi-synthetic data based on real-world user-item interaction datasets. Concretely, we adopt MovieLens 1M (\textbf{ML1M})~\cite{harper2015movielens} and  \textbf{Ciao}~\cite{tang2012mtrust} as base datasets and randomly keep 1,000 users in each dataset. Then, we run the matrix factorization (MF) model~\cite{koren2009matrix} to complete the original datasets to provide the ground truth of user-item relevance. The detailed statistics of the semi-synthetic datasets are shown in Table~\ref{table:datasets}, where we also calculate the Gini Coefficient of the item audience size in each dataset to quantify the inherent audience size imbalance. 

Then, we conduct experiment to simulate the process in Algorithm 1. Concretely, we recommend $K=20$ items to users at each iteration; run the simulation for $T=40,000$ iterations; and retrain the recommendation model after every $L=50$ iterations. To simulate user click behavior, we follow \cite{morik2020controlling} and model the click behavior based on the position bias of $\delta_k=1/log_2(1+k)$ to determine whether user $u$ will examine item $i$ at position $k$. We observe a click only if the user examines and likes the recommended item. All experiments are repeated for 10 times.


\subsection{Evolution of Popularity Bias}
\label{sec:DDS_RQ1}

\begin{figure}[t!]
\vspace{-20pt}
\centering
\includegraphics[ width=0.88\linewidth ]{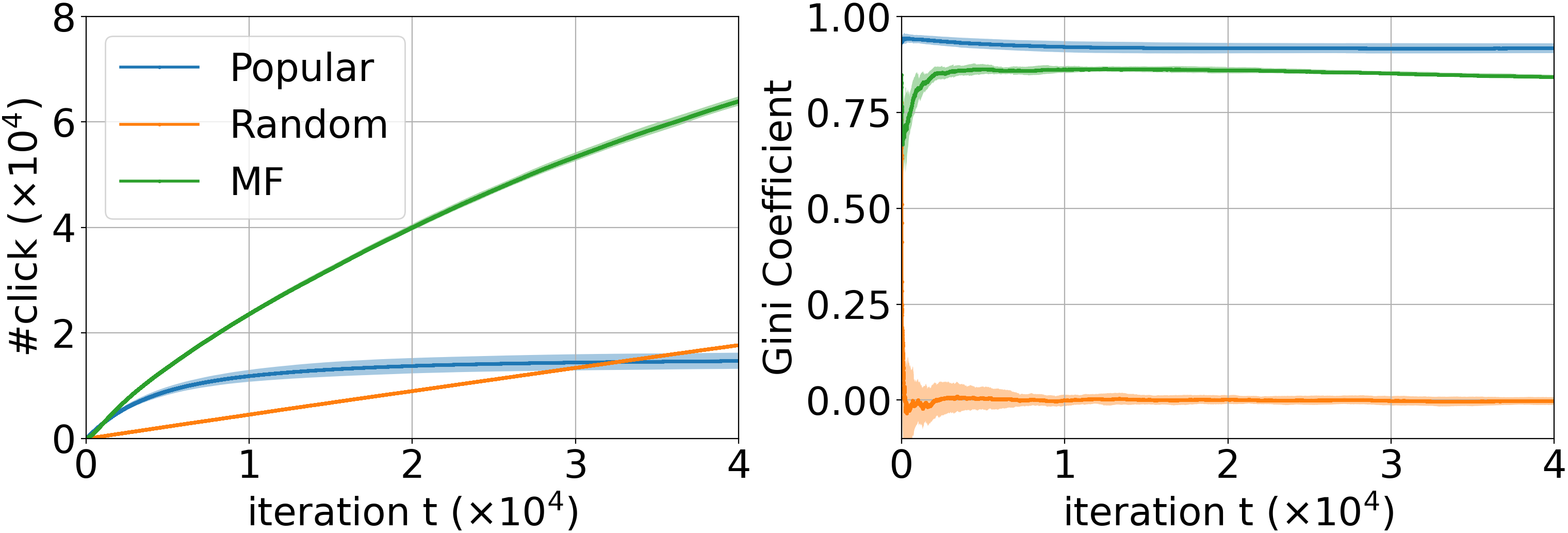} 
\vspace{-8pt} 
\caption{Results of three methods on ML1M.}
\label{fig:DDS_RQ1_ml1m} 
\end{figure}

The first question to investigate is: how does popularity bias evolve in dynamic recommendation? Here, we use the basic MF as the recommendation model~\footnote{In this paper, to counteract the position bias, we adopt the inverse propensity scoring based loss from \cite{saito2020unbiased} for training the MF model, where we use $p_k=1/log_2(1+k)$ as the propensity estimation for a click observed at position $k$}, and the dynamic recommendation process involves all three bias factors introduced in Section~\ref{sec:factors}. Results for ML1M are shown in Figure~\ref{fig:DDS_RQ1_ml1m}, where for comparison, we also include a \textbf{Popular} method to rank items only based on the number of observed clicks so far, and a \textbf{Random} method to randomly rank items. At iteration $t$, we report the number of cumulative clicks up to now as the metric evaluating recommendation utility, and we report $Gini$ for measuring the popularity bias.

First, we observe in the left figure in Figure~\ref{fig:DDS_RQ1_ml1m} that MF produces significantly higher recommendation utilities than the Popular and Random methods. Moreover, the number of cumulative clicks first increases then converges for the Popular method, and after some iterations the Random method can even outperform the Popular method on both datasets, which illustrates the harm of popularity bias. Second, we observe in the right figure that: (i) the Random method produces near zero bias; (ii) the Popular method results in high $Gini$ values throughout the whole experiment; and (iii) MF first produces a rapid increase in $Gini$ and then maintains this high $Gini$ value to the end of the experiment.



While it is not surprising that we observe popularity bias in dynamic recommendation, it is surprising that a traditional MF (which is also the foundation of many more advanced models~\cite{niu2018neural,he2017neural,wang2019neural}) boosts the bias so fast, and the produced bias nearly equals that in a heavily-biased Popular method. Beyond static studies~\cite{zhu2021popularity} of popularity bias that have observed its prevalence, we observe that this bias grows rapidly and maintains at a high level, indicating the need for special interventions to mitigate this issue.

\subsection{Impacts of Three Bias Factors}
\label{sec:DDS_factors}

Next, we investigate the impacts of the three factors introduced in Section~\ref{sec:factors}.




\subsubsection{Impact of Closed Feedback Loop}
\label{sec:DDS_RQ3}

First, we conduct a new experiment that removes the closed feedback loop by: (i) not using the clicks collected from personalized recommendation (by MF) as training data; (ii) after every $L$ personalized recommendation iterations, adding a random recommendation step to generate random rankings to $L$ randomly selected users and collect random-recommendation clicks; and (iii) only using the random-recommendation clicks to train the personalized MF model. In this way, the MF is trained by data purely from random recommendations and will not be influenced by previous personalized recommendation models, i.e., breaking the closed feedback loop. We evaluate the popularity bias only for personalized recommendations by MF. We denote this experiment setup as \textbf{w/o CFL}, and denote the experiment with closed feedback loop (the same as MF in Section~\ref{sec:DDS_RQ1}) as \textbf{w/ CFL}.

\begin{figure}[t!]
\vspace{-20pt}
\centering
\includegraphics[ width=.9\linewidth ]{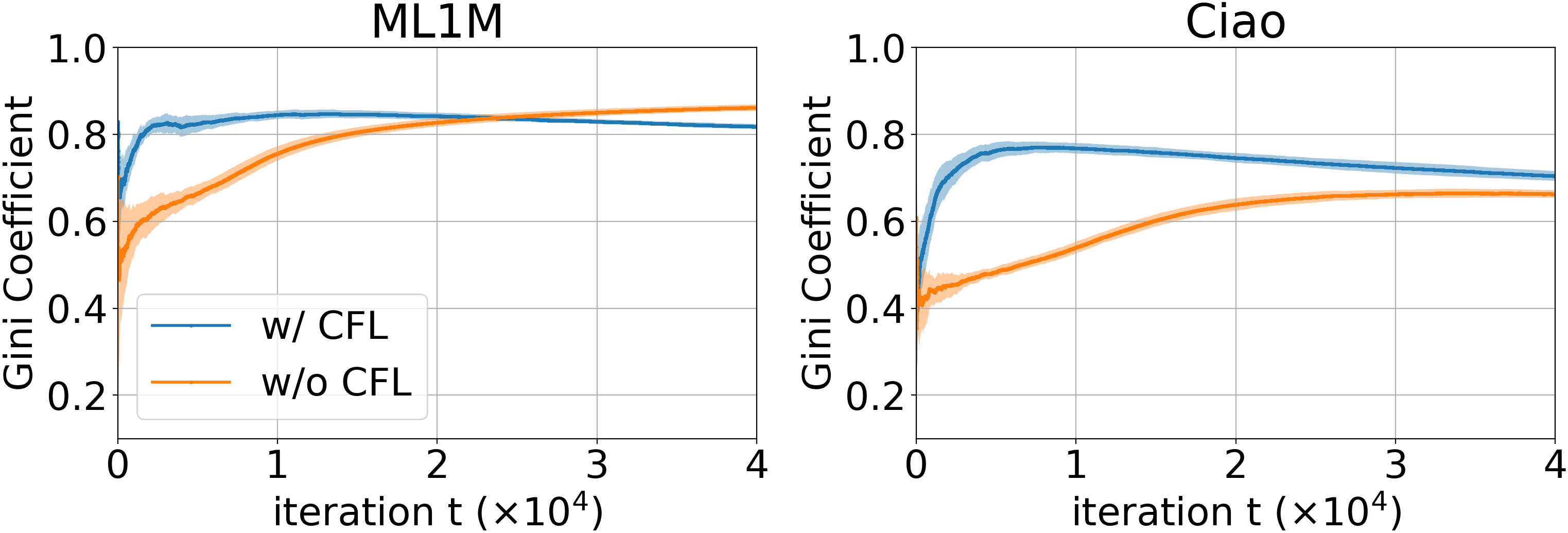} 
\vspace{-8pt} 
\caption{Compare popularity bias in experiments with (w/ CFL) and without closed feedback loop (w/o CFL).}
\label{fig:DDS_RQ3} 
\end{figure}

Because the MF in w/o CFL is trained by click data from random recommendations, whose data size is much smaller than that in w/ CFL. Hence, it is unfair and not informative to compare the utility between w/ CFL and w/o CFL, and we only show the popularity bias comparison in Figure~\ref{fig:DDS_RQ3}. From the figures we can see that compared to w/ CFL, in w/o CFL, the popularity bias also keeps increasing but at a much slower speed. This indicates that \textit{the closed feedback loop does exacerbate the popularity bias}. Without the closed feedback loop, the popularity bias is only from the current recommendation model, and there is no accumulated bias from previous models. Notice that $Gini$ in w/o CFL still keeps increasing. This is because the training data gets increasingly denser, making the model bias increases as we will justify in the following section.

\subsubsection{Impact of Model Bias}
\label{sec:DDS_RQ4}

\begin{figure}[t!]
\vspace{-20pt}
\centering
\includegraphics[ width=.88\linewidth ]{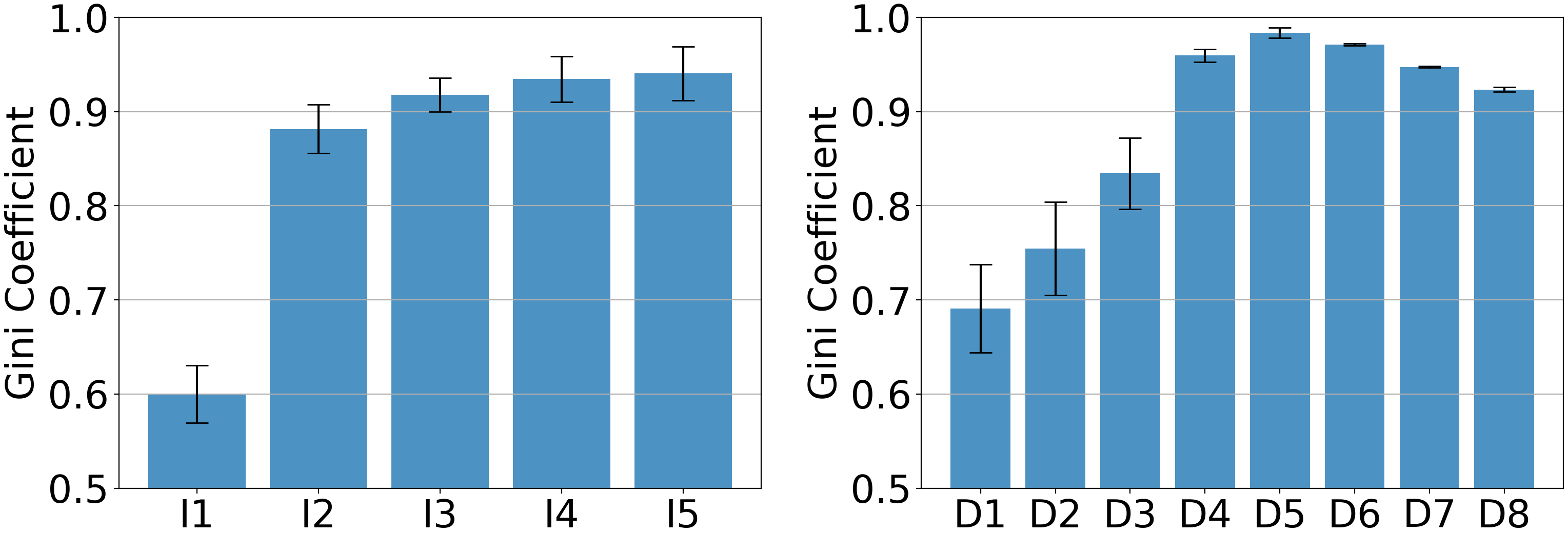} 
\vspace{-8pt} 
\caption{Influence of inherent audience size imbalance (left) and training data density (right) on model bias.}
\label{fig:DDS_RQ4_ml1m} 
\end{figure}

Next, we conduct a series of static recommendation experiments to study the impact of model bias on popularity bias. First, we study how the inherent audience size imbalance influences model bias. Beside the semi-synthetic dataset ML1M we already used, we also generate 4 variants with different levels of inherent audience size imbalance. Now we have 5 datasets with increasing levels of audience size imbalance, denoted as $\text{I}_1, \text{I}_2, \text{I}_3, \text{I}_4, \text{I}_5$, and the corresponding $Gini$ of audience size are 0.37, 0.45, 0.51, 0.57, 0.64 (higher value means severer imbalance). Result of a conventional MF model is shown in the left of Figure~\ref{fig:DDS_RQ4_ml1m}, where we see that severer imbalance leads to increased model bias.

Next, we investigate how training data density influences the model bias. In this case, we use the same ML1M dataset with $Gini$ of audience size 0.64, but generate 8 training datasets with different densities. The 8 training datasets with increasing densities are denoted as $\text{D}_1, \text{D}_2, \text{D}_3, \text{D}_4, \text{D}_5, \text{D}_6, \text{D}_7, \text{D}_8$, and the corresponding densities are 0.01\%, 0.05\%, 0.1\%, 0.2\%, 0.4\%, 0.8\%, 1.6\%, 3.2\%. Experimental result is presented in the right of Figure~\ref{fig:DDS_RQ4_ml1m}, where we can see that with training datasets getting denser, the model bias first increases but then deceases. This may be because with denser data, both model bias and ability to learn user-item relevance are improved. And after a threshold, the ability to learn user-item relevance surpasses the effect of model bias, leading to lower popularity bias observed. However in practice, dense training data is rare and the model bias usually plays a major role.

\subsubsection{Impact of Inherent Audience Size Imbalance}

\begin{figure}[t!]
 \vspace{-20pt}
\centering
\includegraphics[ width=.9\linewidth ]{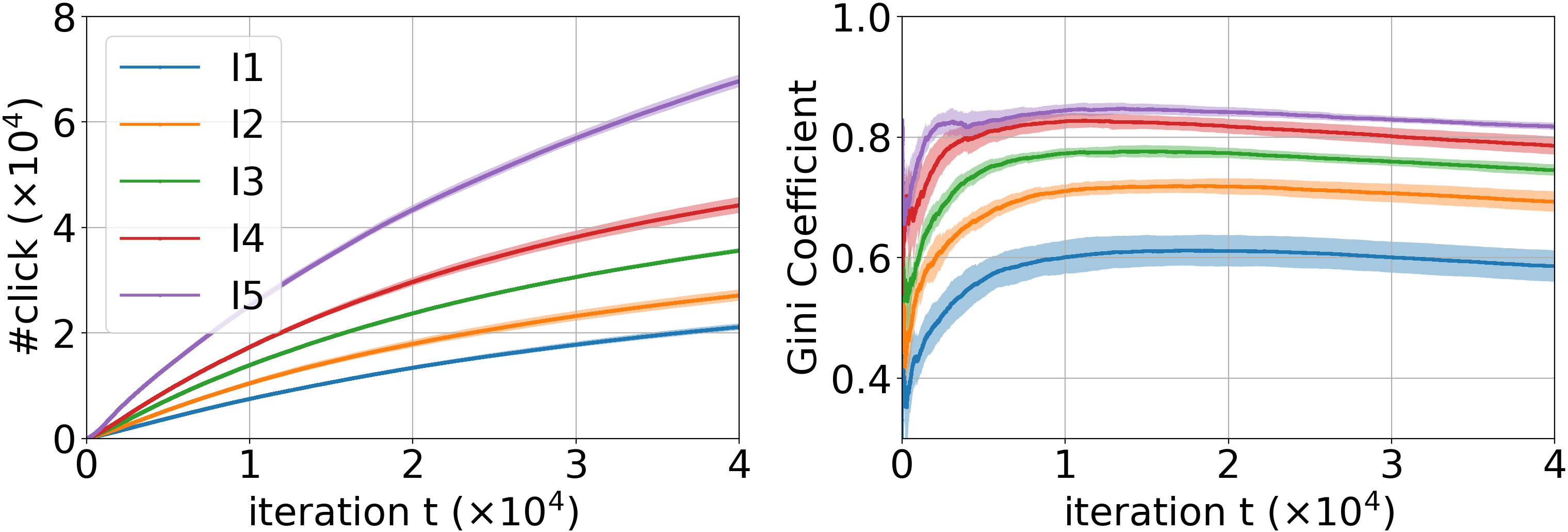} 
\vspace{-8pt} 
\caption{Utility (left) and popularity bias (right) for datasets with different inherent audience size imbalance.}
\label{fig:DDS_RQ5_ml1m} 
\end{figure}

The inherent audience size imbalance exerts its influence on popularity bias mainly through model bias, which we already exhibit by static experiments in Figure~\ref{fig:DDS_RQ4_ml1m}. But how does inherent audience size imbalance impact dynamic recommendation? To answer this, we run dynamic recommendation experiments for the 5 datasets with different levels of inherent audience size imbalance, where all other experiment settings are the same as the MF experiment in Section~\ref{sec:DDS_RQ1}. Results are presented in Figure~\ref{fig:DDS_RQ5_ml1m}. The left figure demonstrates that with severer inherent audience size imbalance, a system can receive more user clicks. This is because popular items can be more easily recognized and correctly recommended to matched users to receive large amounts of clicks in imbalanced datasets. On the other hand, the right part in Figure~\ref{fig:DDS_RQ5_ml1m} shows that with a severer inherent audience size imbalance, higher popularity bias is generated.

\subsubsection{Summary}

In sum, we find that the inherent audience size imbalance and model bias are the main sources of popularity bias; while the closed feedback loop can intensify the bias when inherent audience size imbalance and model bias exist. Moreover, we also find that higher training data density and greater imbalance can increase the effect of model bias.


\section{Debiasing Approaches}
\label{sec:debiasing}

While we have demonstrated the evolution of popularity bias, how can we begin to counteract it? As we empirically studied in Section~\ref{sec:DDS_factors}, model bias is the most essential factor. Thus, in this section, we focus on how to mitigate the popularity bias in dynamic recommendation by reducing model bias.

Most existing works reduce popularity bias in a static setting by reducing model bias~\cite{wei2020model,steck2019collaborative,steck2011item,zhu2021popularity}. For example, \cite{steck2019collaborative} proposes a re-scaling method (denoted as \textbf{Scale}) to reduce the bias by re-scaling the outputs of recommendation models as a post-processing step. Concretely, the re-scaled score for a user-item pair $(u,i)$ is:
\begin{equation}
\begin{aligned}
\centering
\widehat{r}^{(scaled)}_{u,i} = \widehat{r}^{(model)}_{u,i}/(C_i)^\alpha,
\end{aligned}
\label{equ:Scale}
\end{equation}
where $\widehat{r}^{(model)}_{u,i}$ is the output predicted score from a recommendation model; $C_i$ is the number of clicks the item has in training data; $\alpha$ is the hyper-parameter to control the debiasing strength, higher $\alpha$ means more strength for debiasing; and $\widehat{r}^{(scaled)}_{u,i}$ is the re-scaled score used for final ranking.

In static recommendation, this debiasing strength hyper-parameter $\alpha$ is a constant. However, as we see in Section~\ref{sec:DDS_RQ4}, model bias is proportional to training data density and imbalance. Hence, we propose to gradually increase $\alpha$ from 0 with an increasing step $\Delta$ through the dynamic recommendation process. Beyond the specific Scale method~\cite{steck2019collaborative}, most existing popularity debiasing methods~\cite{wei2020model,steck2019collaborative,steck2011item,zhu2021popularity} include such a debiasing strength weight $\alpha$, meaning that we can apply them dynamically in the same way by involving the increasing step $\Delta$.

Besides, we notice that in a high popularity bias case, popular items can be incorrectly over-recommended to unmatched users (generating false positive signals), the false positive signal is correlated with the popularity bias. If we could correct the recommendations based on these false positive signals, we could lower the popularity bias. So, we propose the False Positive Correction (denoted as \textbf{FPC}) method to correct the predicted scores based on false positive signals in a probabilistic way. More specifically, suppose we are going to predict the relevance $\widehat{r}_{u,i}$ between user $u$ and item $i$, and we already have a predicted score $\widehat{r}^{(model)}_{u,i}$ from a model. Assume that item $i$ has been recommended to user $u$ for $F$ times before and has never been clicked, and we record the ranking positions of these $F$ times of recommendation as $\{k_1, k_2, \ldots, k_F\}$. So, the false positive signals can be denoted as $\{c_{k_1}=0,c_{k_2}=0,\ldots,c_{k_F}=0\}$, where $c_k$ represents whether user $u$ clicks the item $i$ ranked at position $k$. We further denote the probability that $u$ likes $i$ as $\theta_{u,i}$; and denote the probability of examining an item at ranking position $k$ as $\delta_k$. Then, we can calculate the conditional probability that $u$ likes $i$ given the false positive signals as:
\begin{equation}
\begin{aligned}
\centering
P(r_{u,i}=1&|c_{k_1}=0,\ldots,c_{k_F}=0)=1-\frac{1-\theta_{u,i}}{\prod^F_{f=1}(1-\delta_{k_f}\theta_{u,i})},
\end{aligned}
\label{equ:FPC}
\end{equation}
where $\theta_{u,i}$ is unknown and needs to be estimated. We can use the prediction $\widehat{r}^{(model)}_{u,i}$ from a model as $\theta_{u,i}$. So, we use Equation~\ref{equ:FPC} with $\theta_{u,i}=\widehat{r}^{(model)}_{u,i}$ and $\delta_{k_f}=1/log_2(1+k_f)$ (as how we model the position bias) to correct model predictions by false positive signals.

Yet, one disadvantage of FPC is that if we use $\widehat{r}^{(model)}_{u,i}$ from a biased model, such as an MF, as $\theta_{u,i}$, FPC is still vulnerable to the model bias. Thus, we propose to use the predictions from a debiased model, such as the Scale in Equation~\ref{equ:Scale} or other debiasing models introduced in Section~\ref{sec:related_work}, to be $\theta_{u,i}$ in Equation~\ref{equ:FPC}. In this case, we can take full advantage of both true positive signals and false positive signals to counteract the popularity bias.

\section{Debiasing Experiments}
\label{sec:Experiments}
In this section, we conduct experiments to show how the popularity bias is mitigated in dynamic recommendation by dynamically reducing model bias and the proposed FPC method.

\subsection{Setup}
The basic setup is the same as Section~\ref{sec:DDS_setup}. To validate our proposed FPC, we include different types of debiasing methods for comparison. The basic recommendation model is the \textbf{MF}. For the debiasing models, first, we consider existing static debiasing method \textbf{Scale}\cite{steck2019collaborative} as introduced in Equation~\ref{equ:Scale}. As introduced in Section~\ref{sec:debiasing}, we also have the dynamic version, denoted as \textbf{DScale}. Last, we have the proposed \textbf{FPC} method to debias based on false positive signals. And we also combine the proposed FPC with DScale to reduce the popularity bias utilizing both true positive and false positive signals, denoted as \textbf{FPC-DScale}. For every debiasing method (except FPC), there is a debiasing strength weight or the increasing step $\Delta$, we tune these hyper-parameters so that all methods achieve similar bias level, and we compare the click counts to compare the performance. Code available is at https://github.com/Zziwei/Popularity-Bias-in-Dynamic-Recommendation.

\subsection{Empirical Results}
In the following experiments, we study the effect of dynamic debiasing compared with static ones; the effect of the proposed FPC; and the effect of integrating FPC with other debiasing methods.

\subsubsection{How do dynamic debiasing methods perform compared with static ones?}

\begin{figure}[t!]
\vspace{-20pt}
\centering
\includegraphics[ width=.9\linewidth ]{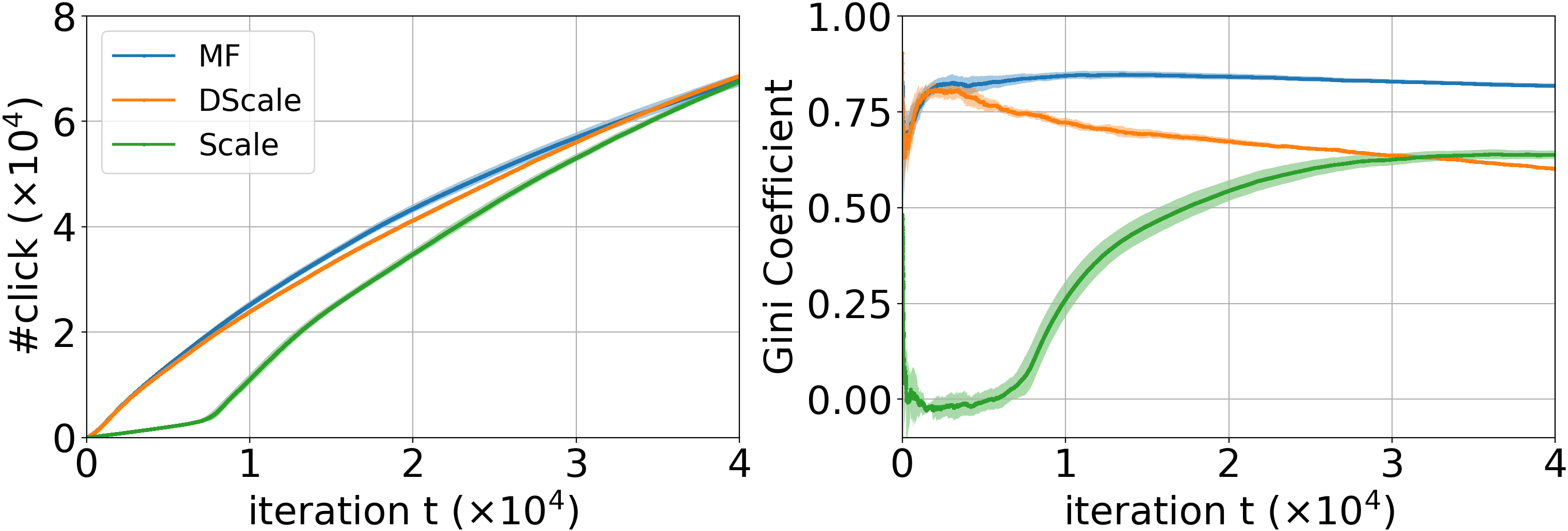} 
\vspace{-8pt} 
\caption{Compare the static debiasing method Scale and its dynamic version DScale.}
\label{fig:static} 
\end{figure}

To show the advantage of dynamic debiasing over static approaches, we conduct experiments with a basic MF and the static debiasing model -- Scale, compared with its dynamic version -- DScale. We tune all models so that similar popularity bias level is achieved, and compare the number of clicks. We plot how the utility and bias change in Figure~\ref{fig:static}. The right figure shows that comparing to MF, both Scale and DScale reduce the bias. However, they show very different patterns: DScale increases the bias at the beginning then keeps decreasing the bias; while Scale keeps increasing the bias and eventually surpasses DScale. This is because as the experiment continues, density and imbalance in training data increases, resulting in higher model bias and more debiasing strength needed. So, dynamically increasing the debiasing strength following the increasing bias can produce better results.

\subsubsection{What is the effect of FPC alone?}

Then, we show the results of MF and FPC in Figure~\ref{fig:FPC_DScale}. The right figure shows that FPC increases the bias at the beginning, but then keeps decreasing the popularity bias. The reduction of bias metric $Gini$ is significant. On the other hand, the left figure shows that FPC can even increase the number of clicks during the experiment compared with MF. This is because by mitigating the popularity bias, popular items are prevented to be over-recommended to unmatched users and more unpopular items are recommended accurately and receive clicks. So, it is a win-win scenario that both users and item providers can benefit from.

\subsubsection{What is the effect of integrating FPC with other debiasing methods?}
\begin{figure}[t!]
\vspace{-20pt}
\centering
\includegraphics[ width=.9\linewidth ]{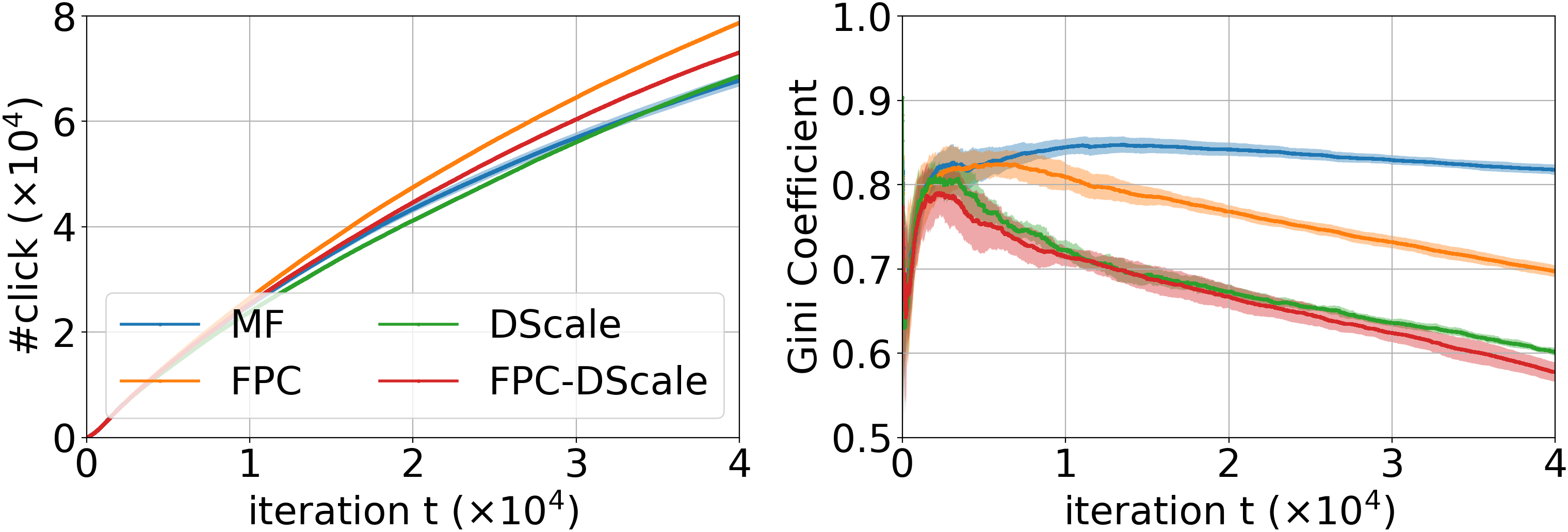} 
\vspace{-8pt} 
\caption{Compare MF, FPC, DScale, and FPC-DScale on ML1M. (Medium debiasing level)}
\label{fig:FPC_DScale} 
\end{figure}

Although, FPC can reduce the bias and improve the utility, it only utilizes the false positive signals without considering the true positive signals as existing debiasing models do. Hence, combining FPC with other debiasing methods is expected to achieve even better performance. To justify this, we also include results of DScale and FPC-DScale in Figure~\ref{fig:FPC_DScale} for comparison. The right figure demonstrates that DScale and FPC-DScale are able to reduce the bias lower than FPC. And we see that DScale and FPC-DScale produce similar level of bias, however, the left figure shows that FPC-DScale generates significantly more clicks, illustrating the advantage of integrating FPC with DScale.

\section{Conclusion}

In this work, we investigate popularity bias in dynamic recommendation. We first conduct an empirical study by simulation experiments to show how the bias evolves in the dynamic process and the impacts of four bias factors on the bias. Then, we propose to dynamically debias and also propose the FPC method to debias utilizing false positive signals. Last, by extensive experiments, we empirically validate the effectiveness of the proposed dynamic debiasing strategy and the proposed FPC algorithm. 




\bibliographystyle{ACM-Reference-Format}
\bibliography{sample-bibliography}

\end{document}